\documentclass[reprint,aps,prf,showpacs,floatfix,notitlepage,superscriptaddress]{revtex4-2}
\usepackage{amssymb,amsmath,bm,bbm}
\usepackage{graphicx}
\usepackage[colorlinks=true,allcolors=blue]{hyperref}
\usepackage{comment}
\usepackage{xcolor}
\usepackage{mathrsfs}

\begin{document}

\title{Topological chiral random walker}

\author{Saeed Osat}
\email{saeedosat13@gmail.com}
\affiliation{Institute for Theoretical Physics IV, University of Stuttgart, Heisenbergstraße 3, 70569 Stuttgart, Germany}

\author{Ellen Meyberg} %
\affiliation{Institute for Theoretical Physics IV, University of Stuttgart, Heisenbergstraße 3, 70569 Stuttgart, Germany}

\author{Jakob Metson} %
\affiliation{Max Planck Institute for Dynamics and Self-Organization (MPI-DS), 37077 G\"ottingen, Germany}

\author{Thomas Speck} %
\email{thomas.speck@itp4.uni-stuttgart.de}
\affiliation{Institute for Theoretical Physics IV, University of Stuttgart, Heisenbergstraße 3, 70569 Stuttgart, Germany}

\begin{abstract}
    Understanding how biological and synthetic systems achieve robust function in noisy environments remains a fundamental challenge across the physical and life sciences. To connect robust behavior with non-trivial topological features present already in the dynamics of individual units, here we introduce the topological chiral random walker (TCRW) model. While exploring the system, a TCRW locates edges and boundaries in the system and develops topologically protected edge currents even in the presence of defects and disorder. Drawing on the bulk-boundary correspondence found in hard condensed matter systems allows us to rationalize the emergence of robust edge currents through topological features of the dynamic spectrum. We show that chiral motion and rotational noise with opposite chirality are two crucial components in our inherently non-Hermitian model. As proofs of principle, we first show that a topological walker outperforms diffusive motion to efficiently solve complex mazes due to its property of remaining on the edge with some rare detachments. Second, we use this model to design building blocks that can perform efficient self-assembly overcoming the timescale bottlenecks of diffusion-limited growth and reducing self-assembly times by approximately 80\%.
\end{abstract}
                           
\maketitle


The concept of topological phases was first suggested in order to unravel the mystery of quantized Hall plateaus in the quantum Hall effect, and it gained widespread attention with the discovery of topological insulators~\cite{Hasan2010RMP}. The need to explain and/or encode robustness in physical systems led topological concepts into other branches of physics including photonics~\cite{Wang2009, Ozawa2019, Lu2014}, meta-materials~\cite{Ni2023}, stochastic systems~\cite{Knebel2020PRL, TangPRX2021, Sawada2024PRL, Jaime2025PPP} and biological and synthetic active matter~\cite{NashPNAS2015, Kotwal2021PNAS, Paulose2015, Souslov2017, Shankar2022NRP, Sone2024review}. A hallmark of a topological phase is topologically protected modes at the boundary of the system. This so called bulk-boundary correspondence suggests that non-trivial topological properties of the bulk give rise to robust gap-less edge modes on the boundary of the system~\cite{Chiu2016RMP, Roman2016PNAS}. These modes are topologically protected and introducing noise and defects in the system will not perturb them as long as the symmetries of the system are respected. 

The field of active matter encompasses systems in which individual components continuously consume energy from their surrounding and convert it into directed motion, leading to collective behaviors that are fundamentally out of equilibrium~\cite{active_matter_roadmap, annurev_active_mat}. Examples of active matter range from microorganisms and cellular tissues to synthetically manufactured particles~\cite{Needleman2017, Marchetti2013RMP}. Recently, the role of ``activity'' to achieve robust function from inherently stochastic dynamics has gained interest, i.e., systems whose qualitative behavior is insensitive to changes of a parameter~\cite{Zheng2024NatComm, Nelson2025, Murugan2017NatComm}. However, many of the topological phenomena observed in active (and passive) systems are limited to emergent features at a coarse level, where the collective behavior of the constituent units gives rise to such phenomena. The prime example is topological defects and their dynamics and manipulation~\cite{Giomi2013PRL, giomi2014defect, Keber2014, Gareth2012RMP, Doostmohammadi2018, Saw2017, Sciortino2023, Shankar2024PNAS,osat2024PRL,Rana2024PRL, Avni2025PRL, Rouzaire2025PRL, Loos2023PRL}. Beyond the approaches based on tracking and manipulating defects and field singularities~\cite{Marchetti2013RMP, BowickPRX2022, Shankar2022NRP}, a different approach to investigate topological phenomena is employing band theory--a systematic approach in condensed matter physics--to define topological phases and corresponding robust edge modes~\cite{Sone2020NatComm, Sone2019PRL, TangPRX2021, Murugan2017NatComm, Dasbiswas2018PNAS, Zheng2024NatComm}. To close the gap between single units and their emerging collective behavior, here we introduce the topological chiral random walker (TCRW) model, a minimal  active matter system with constituent units that show non-trivial topological features already at the single unit level. We show that such a model and its modified versions find applications in different domains.

\section*{Results and discussion}

\begin{figure}[t]
    \includegraphics[width=1\linewidth]{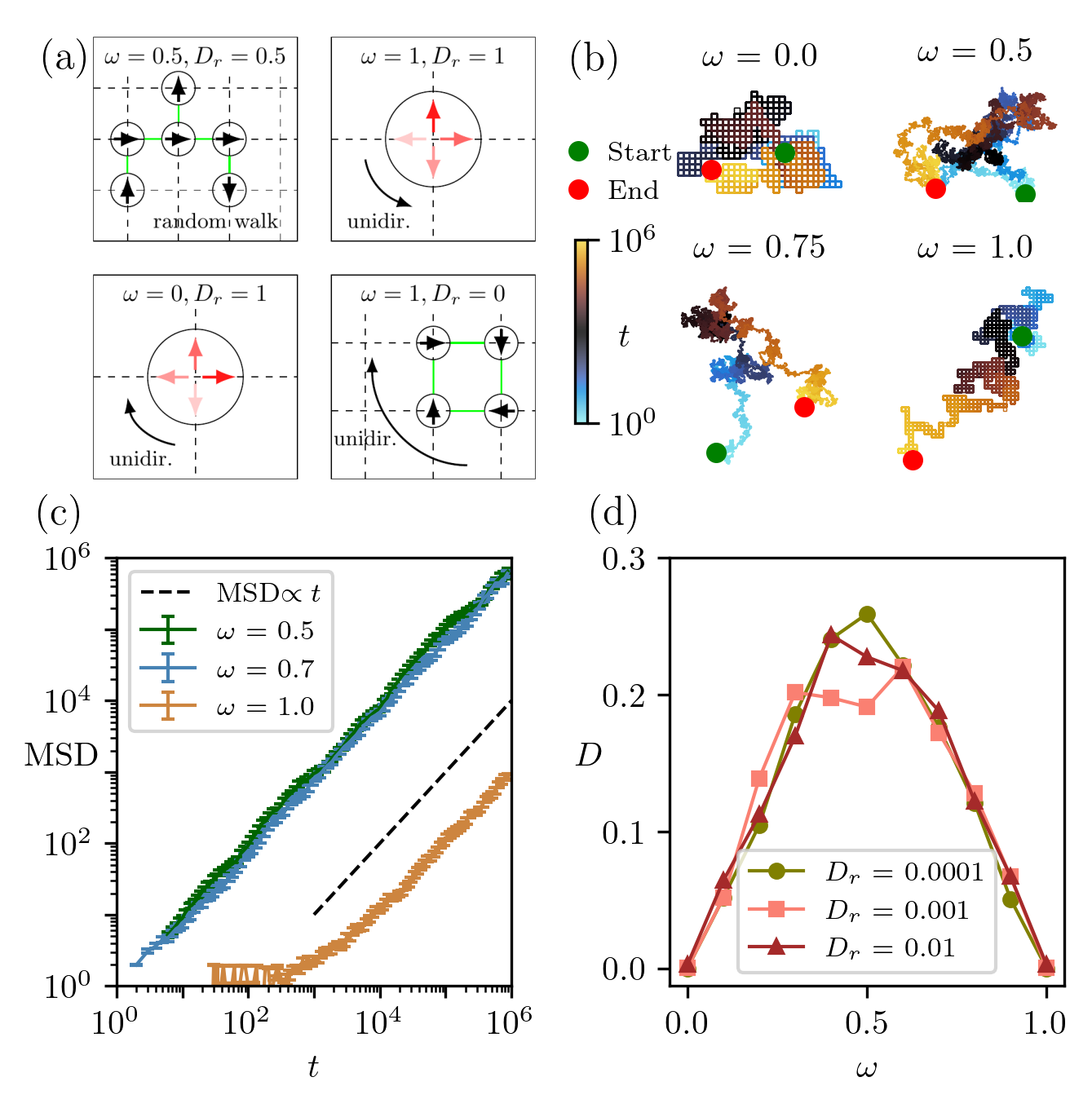}
    \caption{{\bf Chiral random walker.} {\bf(a)} Schematic showing the discrete dynamics of a chiral random walker combining a chiral move (translation and rotation) and rotational noise with opposite chirality. Notice the dynamics of the TCRW for specific values of $D_r$ and $\omega$. {\bf(b)} Sample trajectories of such a walker for different values of chirality~$\omega$ and $D_r=10^{-3}$. {\bf(c)} MSD of walkers for different values of $\omega$ and $D_r=10^{-3}$, confirms that walker experiences normal diffusion. {\bf(d)} The diffusion coefficient decreases linearly with chirality $\omega$ independent of the value of $D_r$.}
    \label{fig:PBC}
\end{figure}

\begin{figure*}[t]
    \includegraphics[width=1\textwidth]{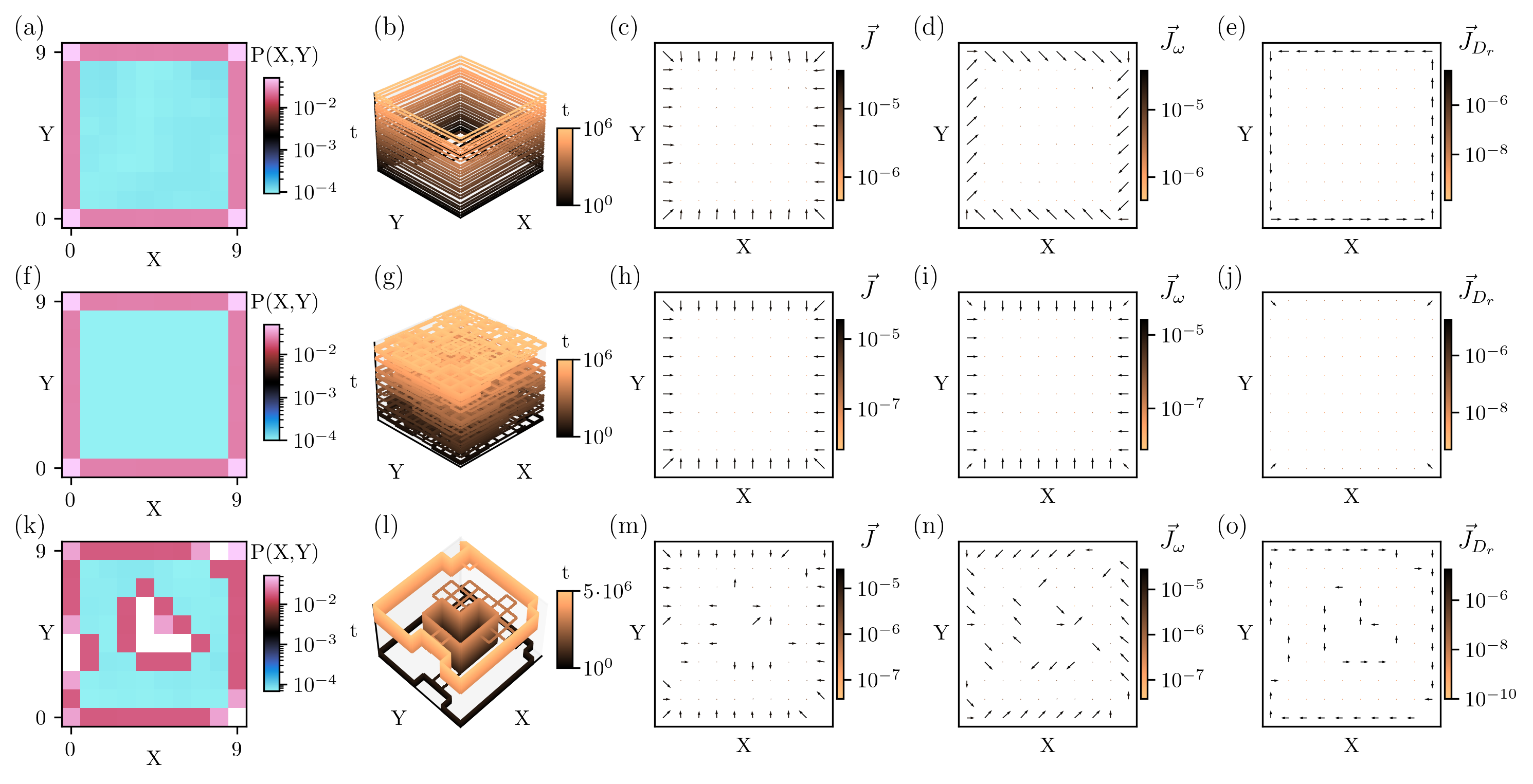}
    \caption{{\bf Chiral edge current.} {\bf(a)} A CRW is placed on a 2D grid with hard walls, where $P(X,Y)$ is the probability of finding a walker on the grid point $(X,Y)$. {\bf(b)} Sample trajectory of the walker for the first $10^6$ steps of the simulation. {\bf(c)}-{\bf(e)} Total current $\bm J$, the chiral current $\bm J_{\omega}$, and the noise current $\bm J_{D_r}$ for $\omega = 1$. {\bf(f)}-{\bf(j)} Corresponding plots for an achiral walker with $\omega=0.5$ and {\bf(k)}-{\bf(o)} for a CRW with $\omega=0$ on a grid with defects on the edge and in the bulk of the system. {\bf (l)} The trajectory of the walker develops edge currents along external and internal boundaries. {\bf (o)} The internal and external boundary edge currents have opposite chiralities. Note that introducing defects and observing the edge current in {\bf (o)} confirms that the system is sensitive to edges due to the topological properties of the dynamics. For simulations we used $D_r=10^{-3}$ and total number of steps $T=10^{10}$.}
    \label{fig:OBC}
\end{figure*}

\paragraph*{Topological chiral random walker.}
Let us consider a chiral random walker (CRW)~\cite{Hargus2021, Sevilla2016, Mallikarjun2023} with an internal degree of freedom, the director $\bm d \in \{ \uparrow, \downarrow, \rightarrow, \leftarrow \}$ which shows the direction of the next step of the walker on a discrete 2D grid. The state of the walker located at lattice point~$(i,j)$ is captured by $(i,j,\bm d)$. At each discrete time step, the walker undergoes a rotational noise step with probability~$D_r$, or makes a chiral move with probability~$1-D_r$. By undergoing rotational noise, the walker does not move and just rotates its director counterclockwise (clockwise) with probability $\omega$ ($1-\omega$). In the chiral move, the walker moves in the direction of $\bm d$ and then rotates its director clockwise (counterclockwise) with probability $\omega$ ($1-\omega$). Note that the combined translation and rotation in the chiral step is considered as one step and if translation is not possible (for example in the presence of a defect) the rotation does not take place. The key component of the model is that rotations in the chiral steps and noise steps have different chiralities. For example $\omega=1$ is a clockwise chiral motion with counter-clockwise rotational noise. Note that extreme cases of $D_r=0$ (and let's assume $\omega=1$) corresponds to a walker with state $(i,j,\uparrow)$ to be confined to a chiral orbit: $(i,j,\uparrow) \to (i,j+1,\rightarrow) \to (i+1,j+1,\downarrow) \to (i+1,j,\leftarrow) \to (i,j,\uparrow)$ as shown in Fig.~\ref{fig:PBC}{\bf(a)}. For a non-zero $D_r$ the walker extends its chiral motions to other lattice neighbors. The other extreme $D_r=1$ (and $\omega=1$) is just a spinor $(i,j,\uparrow) \to (i,j,\leftarrow) \to (i,j,\downarrow) \to (i,j,\rightarrow) \to (i,j,\uparrow)$ where the walker is localized on the lattice point~$(i,j)$ with its director doing counter-clockwise rotation as shown in Fig.~\ref{fig:PBC}{\bf(a)}. 

\begin{figure*}[t]
    \includegraphics[width=1\textwidth]{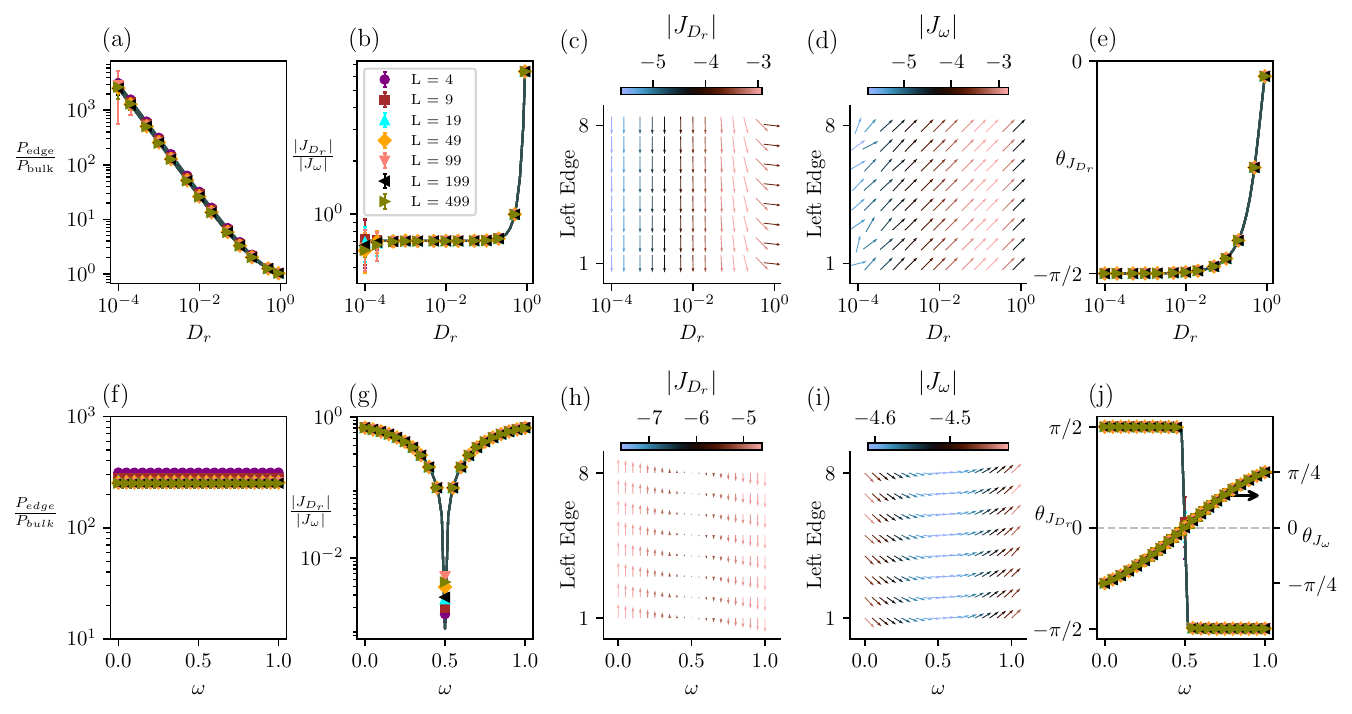}
    \caption{{\bf Impact of rotational noise $D_r$ and chirality~$\omega$.} The black lines show the results obtained from the exact steady-state solution of the transition matrix. {\bf(a)} The ratio $P_{\text{edge}}$/$P_{\text{bulk}}$ becomes independent of system size and confirms the decay of edge localization in the system by increasing $D_r$ for $\omega=1$. {\bf(b)} The ratio of the total values of $\bm J_{D_r}$ and $\bm J_\omega$ are depicted for different $D_r$. The current scatters to the bulk for values of $D_r$ around one. {\bf(c)}-{\bf(d)} shows the orientation of the currents along the left edge for different $D_r$. Note, that the chiral current $\bm J_{\omega}$ is consistently directed towards bulk with $\pi/4$. {\bf(e)} The angle of the current $J_{D_r}$ along the left edge goes from $-\pi/2$ to zero as we vary $D_r$ from zero to one. The angle is measured with respect to the horizontal axis. Note that for all plots in the second row {\bf(f)}-{\bf(j)} there is a symmetry with respect to $\omega = 0.5$. {\bf(f)} The ratio $P_{\text{edge}}$/$P_{\text{bulk}}$ is independent of $\omega$. {\bf(g)}-{\bf(i)} The same as {\bf(b)}-{\bf(d)} but for varying $\omega$. {\bf(j)} The angle of the currents along the left edge with respect to $\omega$.}
    \label{fig:noise_chirality}
\end{figure*}

\paragraph*{Periodic boundary conditions (PBC).}
On a lattice with PBC, depending on the parameters $ 0 \le D_r \le 1$ and $ 0 \le \omega \le 1$ a walker undergoes different dynamics, a) spinor: for $D_r=1$ and any $\omega$ the walker will be localized on a fixed lattice point $(i,j)$ (initial position) and its director undergoes rotational diffusion with chirality~$\omega$, b) chiral rotator: for $D_r=0$ and $\omega=0$ ($\omega=1$) a walker rotates and hops counterclockwise (clockwise) on four adjacent points on the grid, and c) diffusion: for any other values of $(D_r$, $\omega)$ it will diffuse on the grid. For illustration of trajectories of the walker for different chiralities $\omega$, see Fig.~\ref{fig:PBC} \textbf{(b)}. The latter is confirmed by mean square displacement (MSD) of the walker which exhibits a linear behavior as shown in Fig.~\ref{fig:PBC}{\bf(c)}. Note that the diffusion coefficient decreases linearly with chirality Fig.~\ref{fig:PBC}{\bf(d)}. 

\begin{figure*}[!t]
    \includegraphics[width=\textwidth]{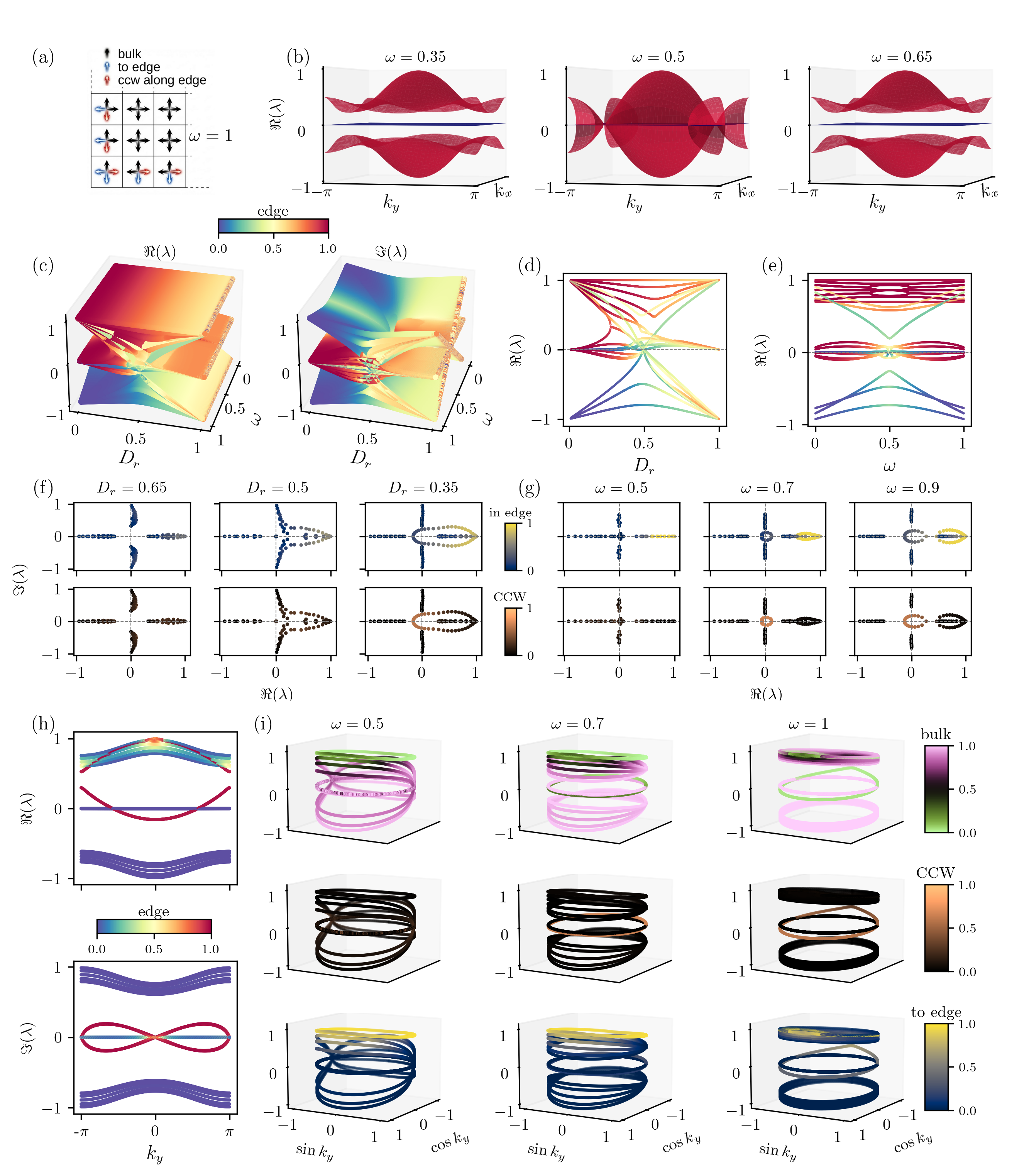}
    \caption{{\bf Topological origin of the edge localization and edge current.} {\bf (a)} Labeling of states for $\omega=1$ where CCW edge current is expected. Edge states are union of ``to edge'' and ``along edge'' states. {\bf (b)} Spectrum of the model in PBC. The gap closes at $\omega=0.5$, where a topological transition occurs. {\bf(c)} Decomposition into real and imaginary part of the spectrum in the presence of OBC for a small lattice of $L=2$. Note that the spectrum is colored based on the localization of corresponding eigenvectors on the edge states. {\bf (d)} Real part of spectrum for a fully chiral case $\omega=1$. Note the coalescence of eigenvalues at $D_r \to 0$. {\bf (e)} Same as {\bf (d)} but for fixed $D_r$. {\bf(f)} Spectrum of the model on a complex plane for a fully chiral case. Coloring differentiates localization on edge and current along edge. {\bf (g)} Same as {\bf (f)} but for changing $\omega$. {\bf (h)} Spectrum of a model with hybrid boundary conditions (periodic along the $y$-direction and open along the $x$-direction). Note the bands localized on the edge. {\bf (i)} Band structure of the HPBC on a closed circle defined by $(\cos k_x, \cos k_y)$. Different coloring is used to show localization on different states.}
    \label{fig:topology}
\end{figure*}

\paragraph*{Open boundary conditions (OBC).} 
In a finite-sized system with open boundaries (breaking translational symmetry) an edge or a boundary is defined as lattice points along the boundary that are not allowed to be occupied by the walker. A walker is placed on a random lattice point in a 2D square grid of size $L\times L$ with OBC and performs a chiral random walk. In both fully chiral $(\omega=0, 1)$ and achiral $(\omega=0.5)$ cases, $P(x,y)$ the probability of finding the walker at the lattice point~$(x,y)$ confirms localization of the walker at the boundary of the system, as shown in Fig.~\ref{fig:OBC}{\bf(a)} and {\bf(f)}. However, comparing the trajectories Fig.~\ref{fig:OBC}{\bf(b)} and {\bf(g)}, one finds that unlike the achiral case, a CRW moves along the edge/boundary in a directed way, an indication of the chiral edge current. To quantify the observed edge current we define the total current $\bm J (x,y)$ as the outflow of the particle from a lattice point~$(x,y)$. As expected, $\bm J$ reflects the fact that walls scatter the walker back to the bulk, see Fig.~\ref{fig:OBC}{\bf(c)}. This suggests that the chiral edge current observed in the trajectories of the walker is hidden in $\bm J$. This is due to the fact that local orbits of chiral moves dominate the edge current. Thus, we decompose $\bm J$ into two components, $\bm J_{D_r}$ and $\bm J_{\omega}$. If a translational move occurs immediately after a noise step then it contributes to $\bm J_{D_r}$, otherwise it is part of $\bm J_{\omega}$. This decomposition correctly captures both the current induced by chiral moves Fig.~\ref{fig:OBC}{\bf(d)} and noise Fig.~\ref{fig:OBC}{\bf(e)}. Note that the edge current runs in the opposite direction to the chirality of the walker, reminiscent of the semi-classical picture of a skipping orbit in the quantum Hall effect. The same decomposition confirms the absence of a chiral edge current for an achiral RW, Fig.~\ref{fig:OBC}{\bf(h)}-{\bf(j)}. Furthermore, the observed edge current for a CRW is robust against defects as shown in Fig.~\ref{fig:OBC} {\bf(k)}-{\bf(o)}. Note that there are two types of defects. Defects that are not introducing new edges (defects that manipulate and deform the current boundary) and internal defects in the bulk of the system which introduce new edges into the system. The walker develops an edge current along both external and internal edges. The direction of the edge current for an internal boundary is the same as the chirality of the walker and opposite to the one for the external boundary as shown in Fig.~\ref{fig:OBC}{\bf(o)}.

\paragraph*{Rotational noise $D_r$ and chirality $\omega$.} For a fully chiral walker $\omega=1$, increasing $D_r$ increases the probability of scattering from the edge to the bulk. To quantify the effect of $D_r$ on the dynamics we decompose $P(X,Y)$ into $P_{\text{edge}}=\sum_{(X,Y) \in \mathrm{edge}}P(X,Y)$ and $P_{\text{bulk}}=\sum_{(X,Y) \in \mathrm{bulk}}P(X,Y)$, the probability of finding a walker on a lattice point on the edge or in the bulk, respectively. As mentioned $P_{\text{edge}}$ decreases by increasing $D_r$ and consequently the reduced probability should appear in the bulk of the system where $P_{\text{bulk}}$ increases with $D_r$ (see Supplemental Material). The localization of the walker is captured by the ratio $P_{\text{edge}}/P_{\text{bulk}}$ which decreases linearly with $D_r$, Fig.~\ref{fig:noise_chirality}{\bf(a)}. To quantify the effect of rotational noise on the current we consider the left edge/wall in the system. The magnitude of the current on the left wall for both $|\bm J_{D_r}|$ and $|\bm J_{\omega}|$ is calculated and their ratio is shown in Fig.~\ref{fig:noise_chirality}{\bf(b)}. The ratio is constant for small values of $D_r$ and starts to diverge for higher values of $D_r$. This seems to contradict the fact that higher $D_r$ values increase the scattering of walker from the wall. However, at higher $D_r$ many of the steps are counted as steps after rotational noise, hence, the ratio $|\bm J_{D_r}|/|\bm J_{\omega}|$ diverges despite the lack of strong edge current. Fig.~\ref{fig:noise_chirality}{\bf(c)} shows the current vector on the left wall for a system of size $L=10$ and $\omega=1$.  Initially $\bm J_{D_r}$ points downwards as expected from the counter-clockwise edge current. Increasing $D_r$ leads to scattering of the current towards the bulk. This is captured by angle of the current vector shown in Fig.~\ref{fig:noise_chirality}{\bf(e)}. Note that the direction of chiral current $|\bm J_{\omega}|$ is constant and equal to $\pi/4$ as expected from clockwise chiral motion (see Extended Data Fig.~\ref{fig:mult_state_sys}{\bf (d)} and Fig.~\ref{fig:noise_chirality}{\bf(d)}). An intuitive description of the mechanism leading to the movement along the edge is given in \emph{Methods}, where we explain how this movement can be mapped onto a dynamics in an effective two-state model.

We perform the same analysis to quantify edge localization and currents for a CRW for a fixed $D_r=10^{-3}$ but varying the chirality $\omega$. Fig.~\ref{fig:noise_chirality}{\bf(f)} shows that edge localization $P_{\text{edge}}/P_{\text{bulk}}$ is independent of chirality. This justifies that solely looking at the probability $P(X,Y)$, one cannot differentiate the topological case when both edge localization and edge currents exist ($\omega=0$ or $\omega=1$) from the trivial case where only edge localization is present ($\omega=0.5$). In contrast to edge localization, the current changes with chirality. $\bm J_{D_r}$ decreases several orders of magnitude when going from a fully chiral $\omega=0(1)$ to an achiral $\omega=0.5$ case. However, $\bm J_{\omega}$ is almost constant (see Supplemental Material). The ratio of the magnitudes of the currents is shown in Fig.~\ref{fig:noise_chirality}{\bf(g)}. A symmetric behavior is observed where the edge current decreases as we span the two extreme CRW chiralities $\omega=0$ and $\omega=1$. The magnitude of $J_{D_r}$ gradually decreases as one moves from $\omega=0$ to $\omega=0.5$ and restores back by moving towards $\omega=1$. At $\omega=0.5$ the walker behaves almost like a normal random walker where at each time step it selects between CW and CCW options with equal probability. For extreme values of $\omega=1$ ($\omega=0$) a CRW undergoes chiral moves which when combined with a small $D_r$ leads to edge currents along the boundaries. Moreover, the direction of $\bm J_{D_r}$ along the left edge of the system shows an interesting behavior where the edge current changes chirality as soon as we cross the achiral regime $\omega=0.5$ as shown in Fig.~\ref{fig:noise_chirality}{\bf(h)}. The edge current which was directed towards $+\pi/2$ at $\omega=0$ decreases in magnitude but keeps its direction until $\omega=0.5$. Then it changes direction to $-\pi/2$ and increases in magnitude until the other extreme value at $\omega=1$ is visited. Note that direction of $\bm J_{\omega}$ shows a continuous change from $-\pi/4$ to $+\pi/4$ upon sweeping the chirality from $\omega=0$ to $\omega=1$. Fig.~\ref{fig:noise_chirality}{\bf(h)} and {\bf(i)} shows the current along the left edge as we change the chirality $\omega$. The average angle of the both currents are shown in Fig.~\ref{fig:noise_chirality}{\bf(j)}. We showed that upon encountering an edge, a walker develops a chiral edge current for a sufficiently weak rotational noise~$D_r \ll 1$ and strong chirality. Note that all the results obtained from simulation are in full agreement with steady-state solution of the transition matrix and also with theoretical results (see Supplemental Material).

\begin{figure*}[t]
    \includegraphics[width=1\linewidth]{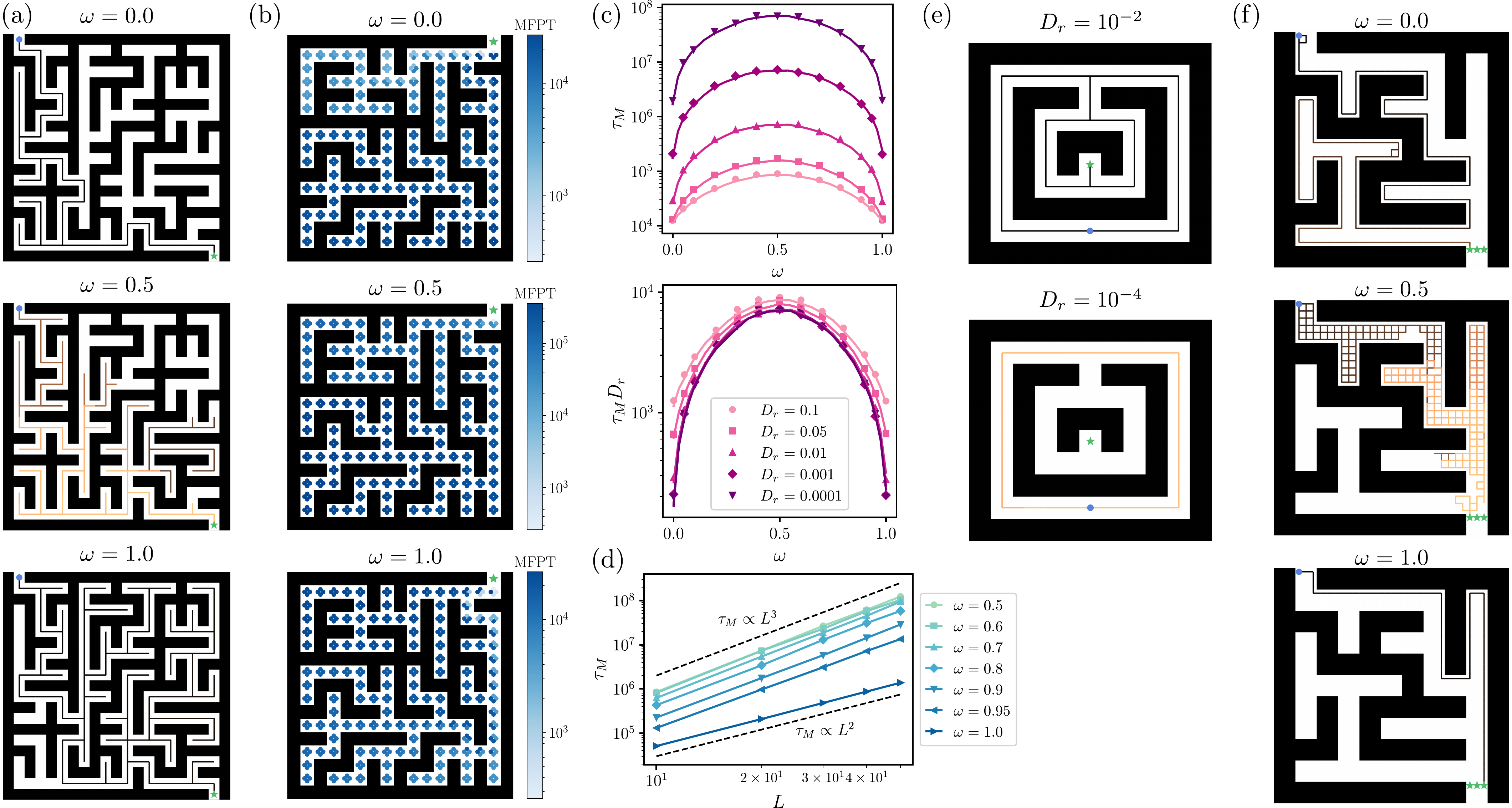}
    \caption{{\bf Maze solving.}
    {\bf(a)} Trajectories of chiral walkers (top and bottom) and an achiral walker (middle). Orange shading indicates the walkers' paths. The blue circle marks the maze entrance, and the green star marks the exit of the maze.
    {\bf(b)} Mean first-passage time $\tau_M$ to solve the maze, depending on the starting position and initial director.
    {\bf(c)} Maze-solving time $\tau_M$ as a function of the chirality $\omega$, for different $D_r$. Top: unscaled. Bottom: rescaled by $D_r$. Lines show the average solving time measured from the transition matrix of a maze, and the markers show the average of times measured directly from simulations of a chiral walker.
    {\bf(d)} Scaling of $\tau_M$ with maze size $L$. The upper dashed line shows $\tau_M\propto L^3$ and the lower dashed line shows $\tau_M\propto L^2$. {\bf(e)} Solving disconnected mazes. {\bf(f)} Solving mazes with wider passage ways ($D_r=10^{-3}$).}
    \label{mazes}
\end{figure*}

\paragraph*{Topological origin of the chiral edge current.}
The dynamics of the walker can be described by a discrete-time Markov chain with transition probabilities organized into a non-Hermitian matrix $P$, see \emph{Methods}. Its general properties guarantee a single real eigenvalue of unity (corresponding to the steady state) with all other eigenvalues lying within the unit disk. The spectrum of $P$ describes the relaxation of both spontaneous and forced excitations towards the steady state. Figure~\ref{fig:topology}{\bf (b)} shows the real part of the spectrum of $P$ (for PBC) in Fourier space, which confirms a gap closure at $\omega=0.5$. The same behavior is observed for a fixed $\omega$ and $D_r=0.5$ (see Fig.~\ref{fig:invariant_1} {\bf (e)}, {\bf (l)}, {\bf (i)}).

To relate the closing of the gap to a topological phase transition, we apply the bulk-boundary correspondence, i.e., the non-trivial topological invariant of the bulk in PBC should lead to a gapless edge mode in OBC. Symmetries of the systems (see \emph{Methods}) suggest deploying the Zak phase as a topological invariant~\cite{Nelson2025, TangPRX2021, Obana2019}. We have calculated the vectorized 2D Zak phase, which confirms the existence of a non-trivial topological phase for $D_r<0.5$ and $\omega \in [0,1]/\ \{0.5\}$ (see Fig.~\ref{fig:invariant_1} {\bf (a)}-{\bf (b)}). A unique feature of non-Hermitian topology is the role played by the boundary. Comparing the spectrum of a system with PBC and OBC (see Fig.~\ref{fig:topology} {\bf (f)}-{\bf (g)}) shows that in the topological region non-Hermiticity leads to a spread of the spectrum in the complex plane and the emergence of a robust edge mode. Fig.~\ref{fig:topology}{\bf (c)} shows the real and imaginary parts of the spectrum of a small system in OBC in the $(\omega, D_r)$ plane. The coloring is based on the localization of the eigenvectors of corresponding eigenvalues on the edge of the system as shown in Fig.~\ref{fig:topology}{\bf (a)}. The real part of the spectrum with a fixed $\omega=1$ (fully chiral) and with a fixed $D_r$ are shown in Fig.~\ref{fig:topology}{\bf (d)} and {\bf (e)}, respectively. Both plots confirm the results obtained from the analysis of the topological invariant of the model. The spreading of the spectrum is illustrated for a system of size $L=10$ for fixed of $\omega$ and $D_r$, see Fig.~\ref{fig:topology} \textbf{(f)} - \textbf{(g)}. Supplemental Fig.~9-10 give additional insights into the difference of the spectra for PBC and OBC.

We also consider the spectrum of a model with hybrid boundary conditions, where the system is periodic along the $y$-direction and open along the $x$-direction. The band structure as a function of $k_y$ is shown in Fig.~\ref{fig:topology}{\bf (h)}. In Supplemental Fig.~9 we show additional properties of the spectrum for these boundary conditions. There are two bands, each localized at one of the edges and with opposite winding. We decompose edge localization as explained before into to-edge, CCW, and bulk components. Figure~\ref{fig:topology}{\bf (i)} shows the contribution for each state (for real part of the spectrum). Note that for $\omega=0.5$ the system exhibits high localization on ``to-edge'' states, which confirms the previous observation in Fig.~\ref{fig:noise_chirality}{\bf (f)} where $\omega$-independent localization of the walker is observed. For higher $\omega$ the weights of eigenvectors that increase the localization in CCW direction increases, i.e., the probability along the edge gets more biased towards the internal state of the walker that points in the direction in which the edge current flows. Moreover, the walker does not discern external edges due to OBC from internal defect in systems with PBC. The spectrum in the complex plane shows the same characteristic loop that is responsible for the edge current, see Extended Data Figs.~\ref{fig:defect_spec} and \ref{fig:nested}.

\paragraph*{Maze solving.}
As a first application of robust edge currents in real space, we show how a chiral walker can efficiently solve mazes. A topological chiral walker finds edges in the system and moves along them with a particular orientation, effectively implementing the ``hand-on-the-wall'' strategy. This maze solving strategy works for any simply connected maze \cite{Niemczyk2020EXC}.  Since the chiral walker moves along the edges with a particular orientation, it is able to systematically traverse the maze in a similar way.

Trajectories of walkers with different chiralities traversing a maze are shown in Fig.~\ref{mazes}{\bf(a)}. Since the the chiral walkers ($\omega=0,1$) move more systematically through the maze, they are able to solve the maze much more efficiently than the achiral random walker ($\omega=0.5$), which tends to go back on itself far more often. This can be seen in the shading of the trajectories, which are colored by time from dark to light.

The mean first-passage time (MFPT) $\tau_M$ to reach a maze's exit from different starting states within the maze is shown in Fig.~\ref{mazes}{\bf(b)} for different values of $\omega$. For the chiral cases, opposite sides of the maze have significantly different solving times, depending on whether it is beneficial to use the left or right hand on the wall. In Fig.~\ref{mazes}{\bf(c)}, we show the MFPT for a walker to reach the exit of 1000 $20\times20$ mazes randomly generated using Prim's algorithm. The markers show the time measured directly from simulations of a chiral walker, and the lines show the MFPT measured from the transition matrix of each maze (see Supplemental Material). The time $\tau_M$ decreases as the chirality $\omega$ moves away from 0.5 (achiral) towards 0 or 1 (fully chiral). As $D_r$ is decreased, the MFPT increases since the walkers require a rotational noise step to move along the wall, which on average requires more time steps for a smaller $D_r$. To account for this effect, in the panel underneath we rescale the MFPT by $1/D_r$. For $D_r \lesssim 0.01$ the curves collapse, indicating that walkers with different $D_r$ are equally efficient if the probability of a noise step is taken into account. For $D_r \gtrsim 0.01$, the walkers become less efficient since for higher $D_r$ values the walkers are more likely to stray from following the wall, leading to unnecessary backtracking and re-exploration, similar to the achiral walker. In Fig.~\ref{mazes}{\bf(d)}, we show how the MFPT to reach the exit scales with the side length $L$ of the maze. We observe a power-law $\tau_M\sim L^3$ for most values of $\omega$. Strikingly, as the walker approaches the fully chiral limit ($\omega\to 0$ or $\omega\to 1$), the scaling decreases to $L^2$ (see Supplemental Material).

The ``hand-on-the-wall'' rule can fail to solve mazes that are not simply connected, as the walker can get stuck on one edge and fail to explore the full maze. An example of this in a disconnected maze is shown in Fig.~\ref{mazes}{\bf(e)}. In the bottom panel of Fig.~\ref{mazes}{\bf(e)} a walker with very small $D_r$ is not able to fully explore the maze within the simulation time, and instead is fixed to traverse the outer edge. Increasing $D_r$, the walker has a higher chance to scatter off an edge. Hence, while the walker does not follow the ``hand-on-the-wall'' rule strictly anymore, it allows the walker to more efficiently solve a disconnected maze as shown in the top panel of Fig.~\ref{mazes}{\bf(e)}.

So far we have looked at situations where the passageway size is equal to the walker size. However, the chiral walker is also able to solve mazes which are wider than the walker itself. In Fig.~\ref{mazes}{\bf(f)} we show such examples, where the thickness of the mazes' walls and passageways is three times the size of the walker. The walkers are able to navigate the maze to the exit, and again by following the edges the chiral walkers are able to find the exit more efficiently than an achiral walker.

While here we have considered a walker solving a physical maze in real space, more abstract state spaces with topological features can be found in many diverse systems. Based on the results of this section, topology could be used to find target states in more general state spaces with complex maze-like structures.

\begin{figure*}[t]
    \includegraphics[width=1\linewidth]{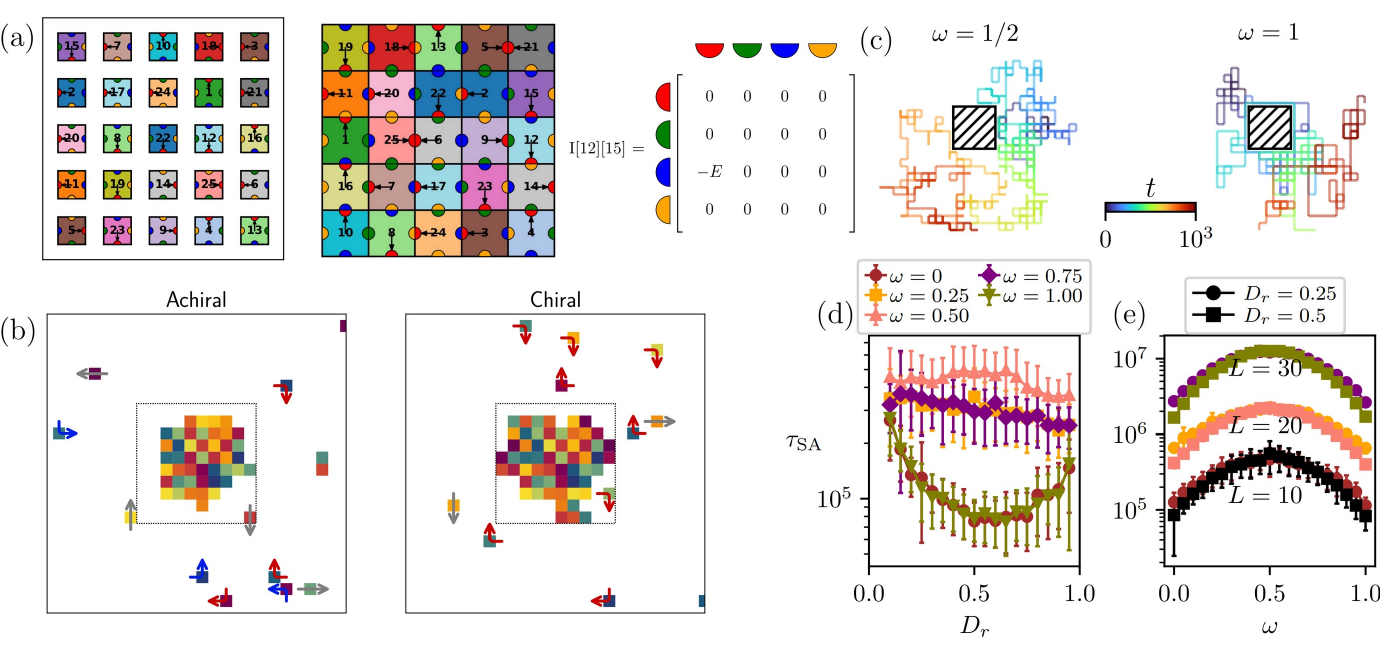}
    \caption{{\bf Self-assembly.} 
    {\bf(a)} Multifarious self-assembly: A set of 25 unique patchy tiles (left) are designed to self-assemble a $5 \times 5$ square target structure (center) deploying patchy interaction between the tiles as encoded in the interaction matrix $I$ (right). {\bf(b)} Tiles undergo random motion for achiral case. Right panel shows that motion of tiles is more directed due to chirality and noise combination (see Supplemental Material). {\bf(c)} Sample trajectories of a tile during SA for achiral ($\omega=0.5$) and chiral ($\omega=1$) cases. The reason for fast self-assembly is due to movement along the edge induced by chirality, which keeps the tile around the initial seed and increases the chance for the tile to find the right place on the initial seed. Note that snapshots are limited to the area close to the initial seed. {\bf(d)} Mean first-passage times $\tau_{\text{SA}}$ for different values of the rotational noise $D_r$. Note that for the fully chiral cases corresponding to $\omega=0$ and $\omega=1$ self-assembly is very efficient. {\bf(e)} Effect of both chirality and system size for a fixed value of rotational noise $D_r$.}
    \label{SA}
\end{figure*}

\paragraph*{Efficient self-assembly.}
As a second application, we show how a modified version of TCRW can tackle the notoriously challenging problem of self-assembly. Self-assembly addresses the problem of building bigger structure from small building blocks~\cite{Whitesides2002S, Glotzer2004S}. Self-assembly is highly relevant both in synthetic and biological systems, where many of the complex proteins and macro-molecules are composed of smaller amino acids or proteins~\cite{Kushner1969, Whitelam2015ARPC, Hormoz2011PNAS, Nguyen2016PNAS, Jacobs2025ARCMP}. 

The process of self-assembly is diffusion limited. A seed or structure only can grow if monomers and building blocks diffuse and reach the growing front. Diffusion-limited growth makes self-assembly inherently inefficient, especially for very dilute systems. Here, we show that tiles designed similar to TCRW can solve this notorious timescale problem. The idea is to use the fact that a CRW finds edges in the system and moves along them. 

We consider a simple case where a square structure of size $L\times L$ is made from $M=L^2$ unique square tiles as shown in Fig.~\ref{SA}{\bf (a)}. The target structure is encoded in the interaction matrix of the tiles using the following Hebbian-like learning rule. Two tiles interact specifically if they are neighbors in the target structure (Fig.~\ref{SA}{\bf (a)})~\cite{Murugan2015PNAS, Sartori2020PNAS, Metson2025JCP-a}.

Now, we endow each tile with a dynamics similar to the CRW introduced above. Each tile undergoes a main chiral move with chirality~$\omega$ which takes place at each time step. On top of the chiral move the tile experiences a rotational noise with opposite chirality with probability $D_r$ (see Supplemental Fig.~2). Note that this modified dynamics is not the same as topological one any more but it enables the tiles to undergo directed motion and also move along the edges in the system Fig.~\ref{SA}{\bf (b)}.

Fig.~\ref{SA}{\bf(c)} shows sample trajectories of one tile as self-assembly begins. In the achiral case ($\omega=0.5$) the dynamics of the tile is similar to a random walk. The tile can reach the boundary of the seed and if not facing a favorable matching tile, will continue the dynamics and diffuse away. However, in the case of chiral tiles $(\omega=0,1)$, after the tile reaches the boundary it has a chance to develop a short trajectory along the seed boundary which enhances the chance of finding the matching partner on the front of the growing seed.

We start self-assembly simulations with an initial small seed (see Supplemental Material for details of the simulation). The average number of the time steps that are required to grow an initial seed to the target structure is defined as mean first passage time to SA~$\tau_{\text{SA}}$. Fig.~\ref{SA}{\bf(d)} shows that for moderate values of $D_r$ and fully chiral case~$\omega=0,1$ self-assembly is very efficient. For a fixed value of $D_r=0.25,0.5$ self-assembly shows the same behavior we observed in the maze solving problem as shown in Fig.~\ref{SA}{\bf(e)}; efficient self-assembly irrespective of system size is accessible in chiral case where $\tau_{\text{SA}}$ is on average 80\% smaller than the achiral case.


\section*{Conclusions}

We have introduced the Topological Chiral Random Walker (TCRW), a minimal model that establishes a new paradigm for encoding topological robustness into the dynamics at the single-unit level. Combining chiral translation with rotational noise of opposite chirality leads to an inherently non-Hermitian system, wherein robust edge currents emerge through a topological phase transition. Our work addresses a critical gap in active matter research: while previous studies have focused on coarse-grained collective behaviors, the TCRW demonstrates that topological protection can be engineered into the underlying degrees of freedom of individual agents. Our theoretical analysis confirms a topological phase transition characterized by a gap closure in the spectrum and a non-trivial vectorized 2D Zak phase. A hallmark of this phase is the bulk-boundary correspondence, which gives rise to chiral edge currents that are topologically protected against defects. These currents persist along both external and internal boundaries with opposite chirality. The minimal model introduced here paves the way for novel strategies in, e.g., the design of resilient synthetic soft materials and autonomous robotic swarms.

\section*{Acknowledgments}

EM and TS acknowledge funding through the Deutsche Forschungsgemeinschaft (DFG, German Research Foundation) within the framework of the collaborative research centers “Multiscale Simulation Methods for Soft-Matter Systems" (TRR 146) under Project No. 233630050. EM thanks the International Max-Planck Research School for Intelligent Systems (IMPRS-IS) for support. JM thanks the Max Planck Institute for Dynamics and Self-Organization.


%


\clearpage

\onecolumngrid 
\section*{Methods}
\twocolumngrid 

\begin{figure*}[h!]
    \includegraphics[width=1\linewidth]{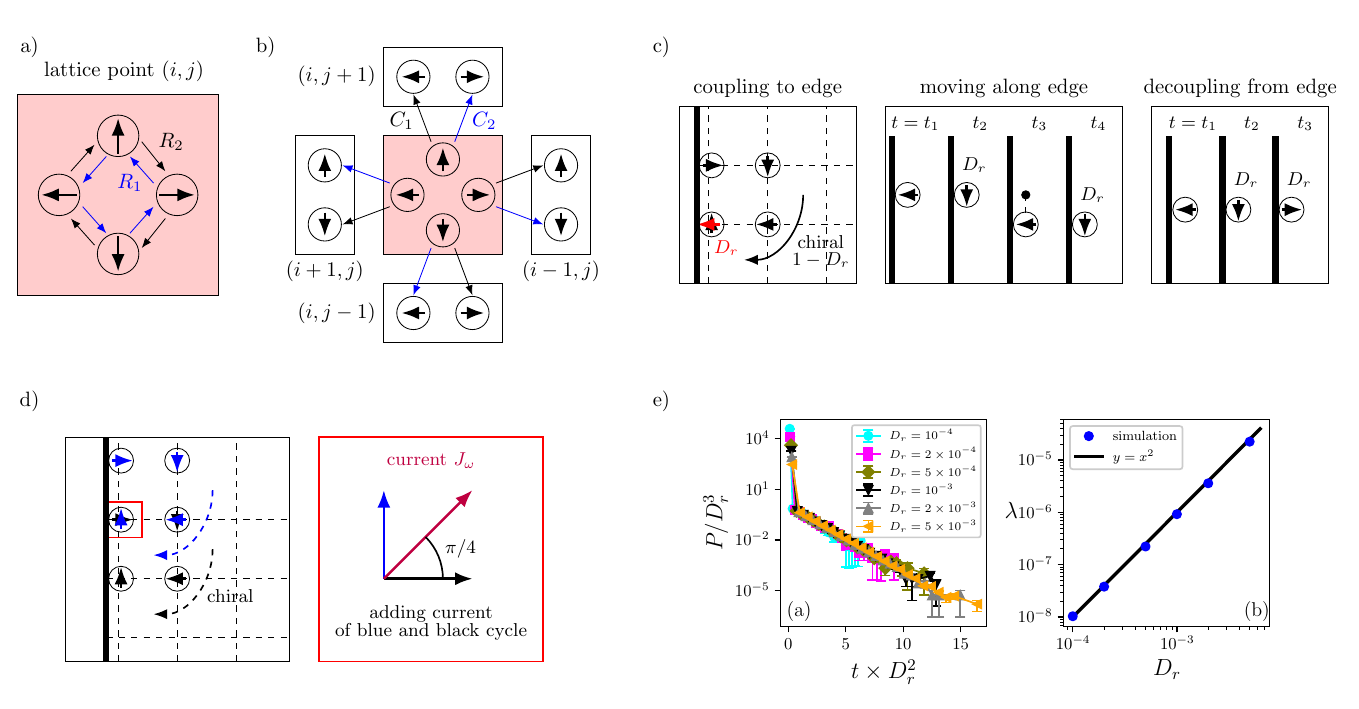}
    \caption{{\bf Sketch of the model.} States $(i,j,\bm d)$ of the walker can be represented by a Markov network. {\bf (a)} Transitions between internal states within a certain lattice side, i.e., orientations of the walker labeled with $s$. Depending on $\omega$ either clockwise or counter clockwise rotations are favored. {\bf (b)} Transitions between neighboring lattice sites labeled with $(i,j)$. {\bf (c)} Coupling and decoupling between walker and edge. The schematics shows the dynamics that (left panel) couples the walker to edge, (middle panel) leads to a movement along the edge and (right panel) leads to a decoupling from the edge. {\bf (d)} Current along the edge. Since the walker may be scattered to the bulk after walking for the edge for a certain time, the current $\bm J_{\omega}$ points towards the bulk. For the fully chiral case $\omega =0,1$ the angle is equal to $\pi/4$. {\bf (e)} Probability distribution of the time spend along the edge. This distribution scales exponentially with  $-D_r^2 $. All curves collapse if they divided by $D_r^3$. }
    \label{fig:mult_state_sys}
\end{figure*}

\begin{figure*}[]
    \includegraphics[width=\textwidth]{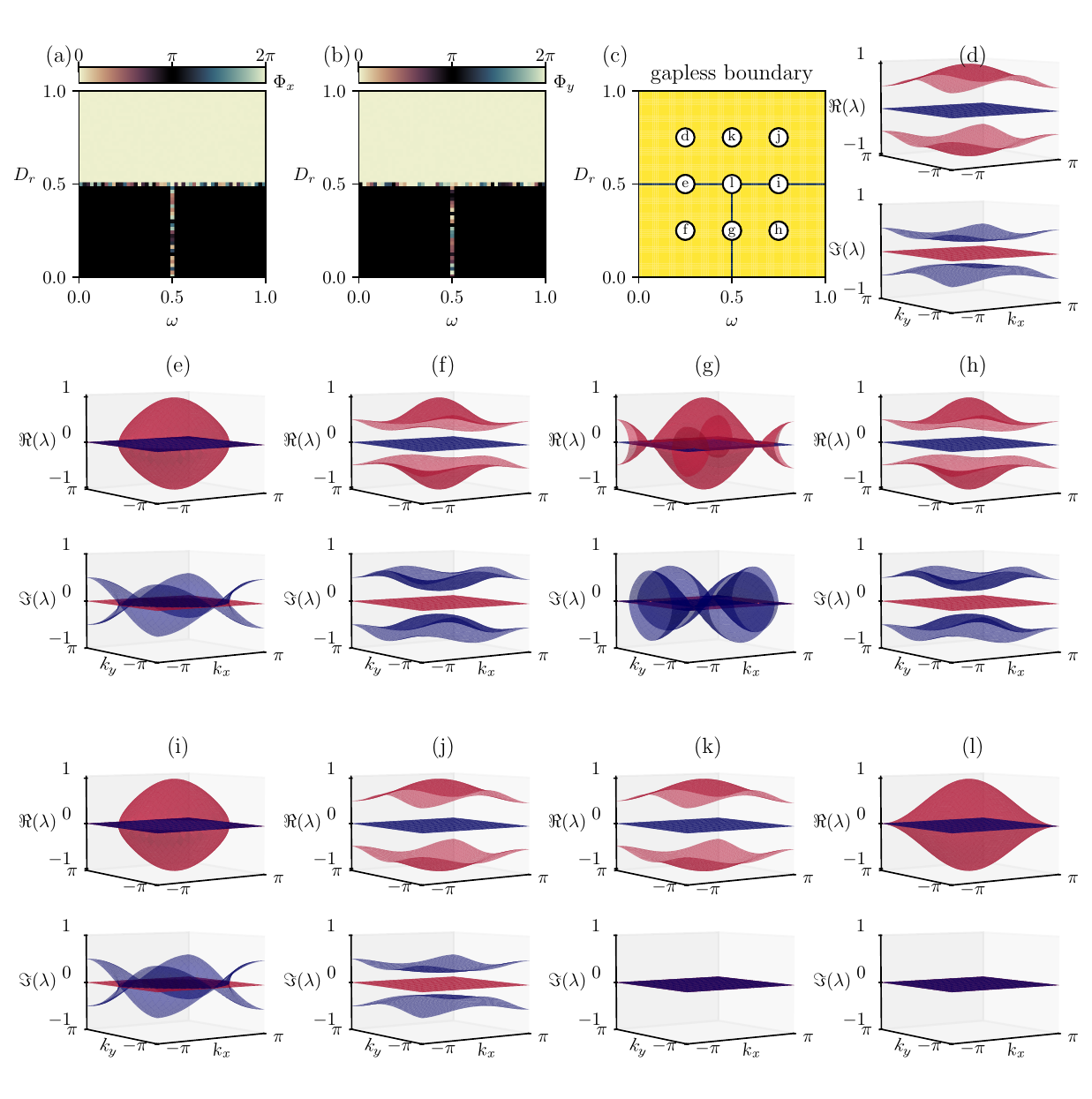}
    \caption{{\bf Topological invariant and gap closing in PBC.} Vectorized 2D Zak phase~$\mathbf{\Phi}=(\Phi_x, \Phi_y)$, topological invariant of the bulk in PBC is calculated using biorthogonal basis of the non-Hermitian transition matrix $W$ {\bf (a)} and {\bf (b)}. Note that the boundaries between topologically trivial and non-trivial phases are in agreement with the boundaries where the gap in the spectrum is closed {\bf (c)}. {\bf (d)}-{\bf(l)} Real and imaginary part of the spectrum in PBC corresponding to 9 points 1-9 as indicated in {\bf (c)}.}
    \label{fig:invariant_1}
\end{figure*}

\paragraph*{Markov chain.}
The evolution of the probability for each state $(i,j,\bm d)$ of the walker is determined by a small set of transition probabilities, which in bulk can be parametrized using four combination
$C_1=(1-\omega)(1-D_r)$, $C_2=\omega(1-D_r)$, $R_1=\omega D_r$, $R_2=(1-\omega)D_r$ since the director on a fixed lattice site possesses an opposite orientation compared to the chiral moves connecting lattice sites, cf. Fig.~\ref{fig:mult_state_sys}{\bf(a)} and {\bf(b)}. For periodic boundary conditions, the transition matrix can be easily expressed in Fourier space
\begin{widetext}
    \begin{equation}
        P(\bm k) = \begin{pmatrix}
            0 & R_1 + C_1 e^{+ik_x} & 0 & R_2 + C_2 e^{-ik_x} \\
            R_2 + C_2 e^{+ik_y} & 0 & R_1 + C_1 e^{-ik_y} & 0 \\
            0 & R_2 + C_2 e^{+ik_x} & 0 & R_1 + C_1 e^{-ik_x} \\
            R_1 + C_1 e^{+ik_y} & 0 & R_2 + C_2 e^{-ik_y} & 0
        \end{pmatrix} 
    \end{equation}
\end{widetext}
with reciprocal lattice vector $\bm k = (k_x, k_y)$. In the Supplemental Material, we solve for the eigenvalues of $P(\bm k)$, which are plotted in Fig.~\ref{fig:invariant_1}. If the walker points towards the edge, chiral moves are forbidden and the walker has to wait for an rotation on the lattice side to occur, i.e., the diagonal entries of $P$ are then $1-D_r$ for each edge state.

\paragraph*{Origin of movement along the boundary.} 
Let us initialize the system with a fully chiral walker exactly on the edge of the system and set its director pointing perpendicular to the edge as shown in middle panel of Fig.~\ref{fig:mult_state_sys}{\bf(c)}. The walker cannot experience any translational move since the path is blocked by the edge. The walker has to wait for a rotational noise step in order to rotate its director. However, such a rotation still keeps the walker coupled to the wall since a chiral step after such a rotational noise will move the walker along the edge and rotate its director, which makes it again directed perpendicular to the edge. Note that this is due to opposite chirality of chiral move and rotational noise. Only two consecutive rotational noise steps can decouple the walker from the edge as explained before and shown in the right panel of Fig.~\ref{fig:mult_state_sys}{\bf(c)}. For such a walker, a step of $D_r$ will couple it to the edge at $t_{\text{cp}}$ and two steps of $D_r$ will decouple it at $t_{\text{decp}}$ and scatter it back to the bulk. One can define the time that the walker spends on the edge as $\tau_{\text{edge}} = t_{\text{decp}} - t_{\text{cp}}$. We observe that the distribution of $P(\tau_{\text{edge}})$ decays exponentially with exponent $D_r^2$ for a wide range of $D_r$ as shown in Fig.~\ref{fig:mult_state_sys}{\bf(e)} . Rescaling $P(\tau_{\text{edge}})$ with $D_r^{-3}$ and $\tau_{\text{edge}}$ with $D_r^2$ collapses all data points for different $D_r$ as shown in Fig.~\ref{fig:mult_state_sys}{\bf(e)}.

\paragraph*{Symmetries of $P$.}
Note that $P$ has inversion and time-reversal symmetries $ I P(\bm k)  I^{-1} =  P(- \bm k)$ as well as $ P(\bm k) =  P^\ast(-\bm k)$ with complex conjugation denoted by the asterisk, and unitary operator $I = \sigma_x \otimes  1$ composed of Pauli matrix $ \sigma_x$ and identity matrix $1$. In addition, the sublattice symmetry $ S P(\bm k) S^{-1} = - P(\bm k)$ with $ S = 1 \otimes  \sigma_z$ and $S ^{-1} = S$ holds true. These symmetries imply that all eigenvalues appear in complex-conjugated pairs.

\paragraph*{Zak phase.} 
The topology of our model is described by the 2D vectorized Zak phase \cite{Liu2017, Obana2019, Lieu2018}. Note that due to non-Hermiticity of our model, we need to consider a biorthogonal basis to obtain the Zak phase. The Berry connection for the $n$-th band is defined as: $\mathbf{A}_n(k_x, k_y) = i \langle \phi_n | \partial_{\mathbf{k}} | \psi_n \rangle$ which is calculated in biorthogonal basis using left and right eigenvectors $\langle \phi_n |$ and $| \psi_n \rangle$, respectively. The vectorized 2D Zak phase
\begin{equation}
    \mathbf{\Phi}_n = \frac{1}{2\pi} \int_{\mathrm{BZ}} \mathbf{A}_n(k_x, k_y) \, dk_x dk_y.
\end{equation}
then follows through integrating. We plot $\mathbf{\Phi}$ in Fig.~\ref{fig:invariant_1} {\bf (a)} and {\bf (b)}. This highlights, that $D_r = 0.5$ represents a phase boundary between the topological ($D_r < 0.5$) and trivial regime ($D_r > 0.5$). Furthermore, Fig.~\ref{fig:invariant_1} {\bf (d)}-{\bf (l)} shows the real and imaginary parts of the eigenvalues of $P$ with respect to the components of $\bm k $ for different regions of the phase diagram as sketched in Fig.~\ref{fig:invariant_1} {\bf (c)}. 
The topological invariant is $\mathbf{\Phi}=(0,0)$ in the trivial phase and $\mathbf{\Phi} = (\pi,\pi)$ in the non-trivial phase as shown in Fig.~\ref{fig:invariant_1}. The topological transition is accompanied by a gap closure in the spectrum and the emergence of a gap-less edge mode.

\begin{figure*}
    \centering
    \includegraphics[width=\linewidth]{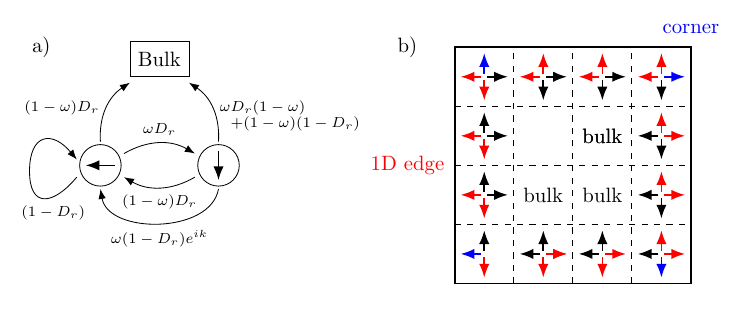}
    \caption{\textbf{Illustration of the one-dimensional movement along the edge.} \textbf{(a)} Two-state model (with one additional absorbing bulk state) that is used to describe the one-dimensional movement along the left edge. \textbf{(b)} Internal states are considered as being on the edge and on the bulk. The corners highlighted in blue are neglected for the effective one-dimensional model. We are thus left with a chain composed of red states that are arranged periodically.}
    \label{fig:1D_model}
\end{figure*}

\begin{figure*}[]
    \includegraphics[width=\textwidth]{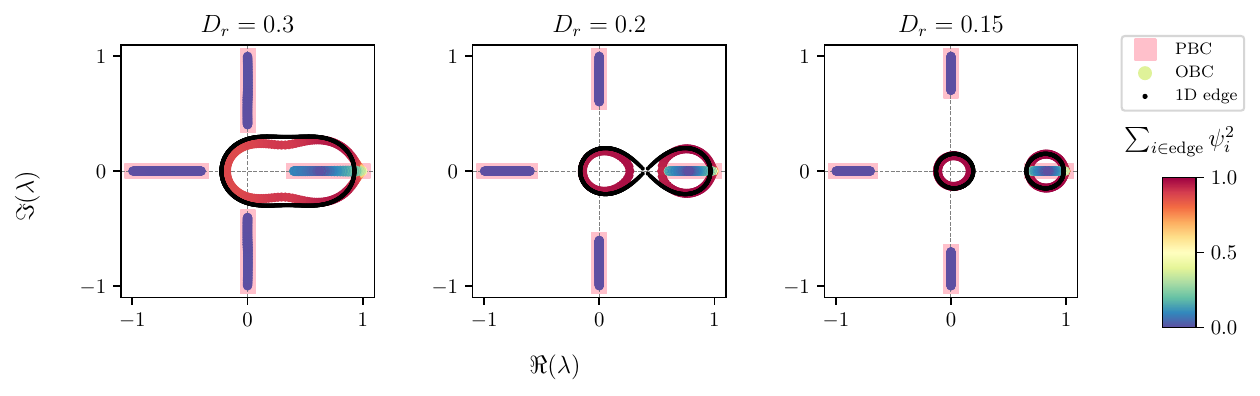}
    \caption{{\bf Time-scale separation of chiral edge current.} In the topological regime ($D_r<0.5$) the oval shape band which is highly localized on the boundary of the 2D square and represents chiral edge current splits in two. This band can be captured by a simple 1D model that mimics the behavior of the boundary, \ref{fig:1D_model}. This splitting is the indicator of time-scale separation in the chiral edge current when long-time hydrodynamic modes close to the steady state (right circle) is separated from short-time transient modes (left circle).}
    \label{fig:time_scale_sep}
\end{figure*}

\begin{figure*}[]
    \includegraphics[width=\textwidth]{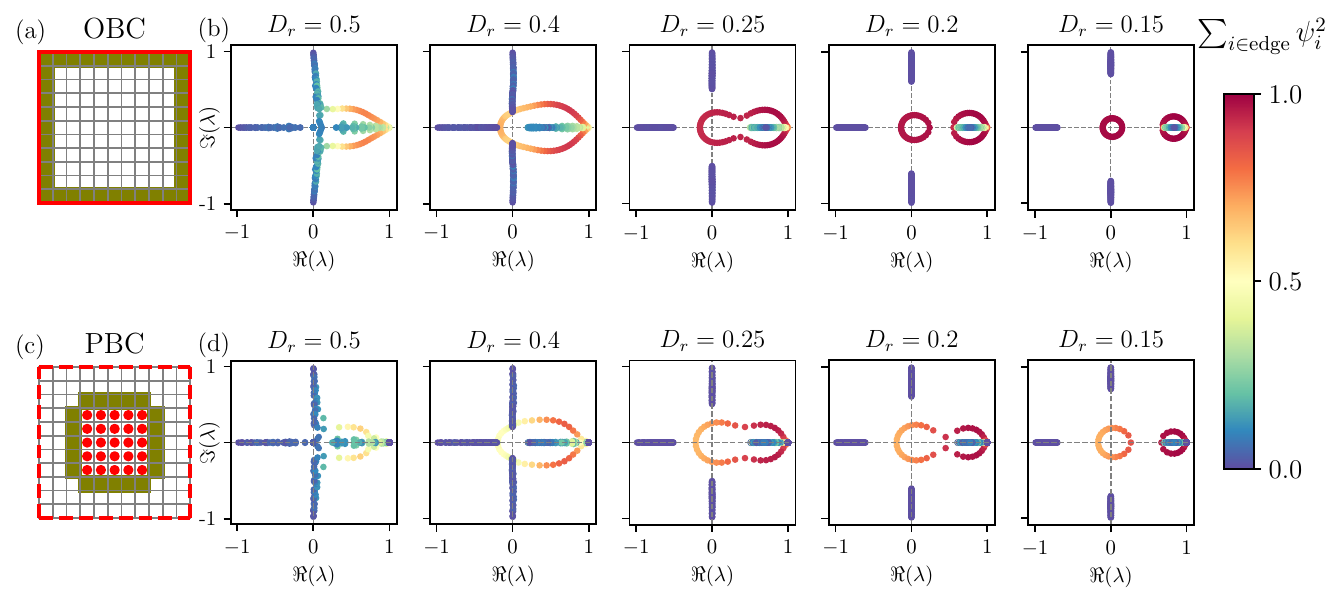}
    \caption{{\bf Effect of boundaries on the spectrum.} {\bf (a)} shows a lattice in OBC. The green color shows the boundary used in coloring localization of spectrum. {\bf (b)} shows the spectrum of {\bf (a)}  for different values of $D_r$ and for constant $\omega=1$. each eigenvalue is colored with its contribution to states that are towards the edge and along the edge (in CCW direction). {\bf (c)} shows a lattice in PBC in the presence of a defect (connected set of red disks). {\bf (d)} is the same as {\bf (b)} but for the internal boundary along the defects. Note that oval like shape in the spectrum originates from the topology and having a boundary in the system.}
    \label{fig:defect_spec}
\end{figure*}

\begin{figure*}[]
    \includegraphics[width=\textwidth]{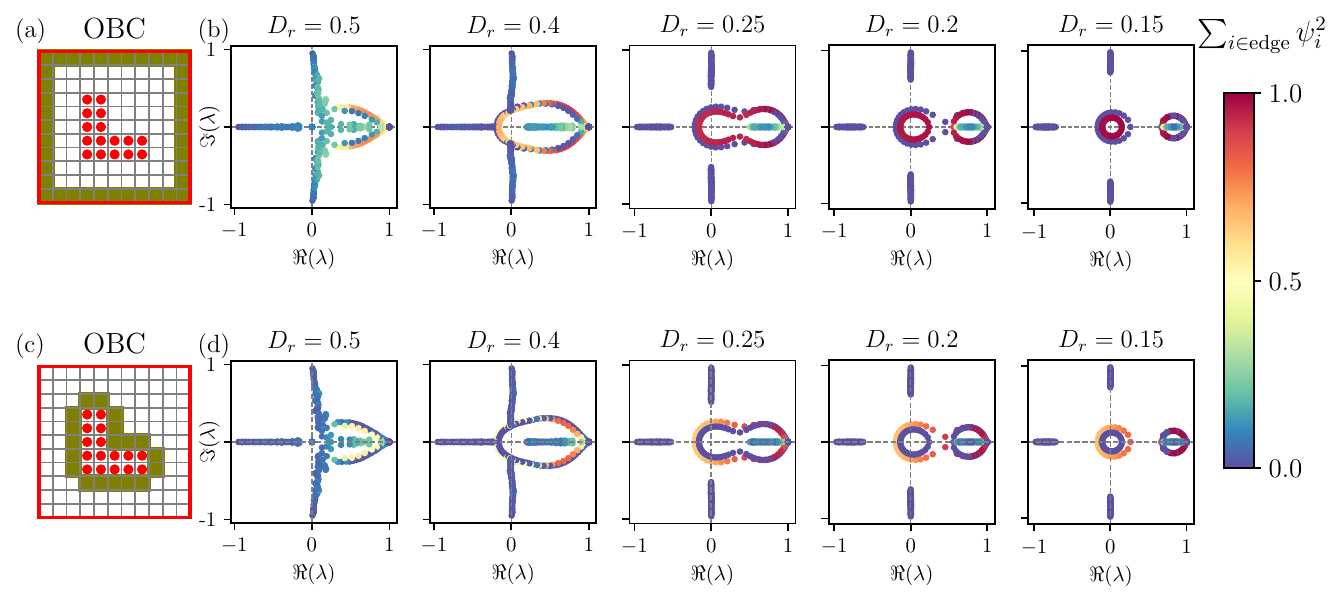}
    \caption{{\bf Disconnected boundaries and nested spectrum.} {\bf (a)} shows a lattice in OBC and in the presence of defects (lattice points with red disks). The green color shows the boundary used in coloring localization of spectrum. {\bf (b)} shows the spectrum of {\bf (a)}  for different values of $D_r$ and for constant $\omega=1$. each eigenvalue is colored with its contribution to states that are towards the edge and along the edge (in CCW direction). {\bf (c)} and {\bf (d)} are the same as {\bf (a)} and {\bf (b)} but for the internal boundary along the defects in CW direction. Note that nested oval like shape in the spectrum originates from the two separate boundaries in the system where the inner one is localized on the external boundary (longest) and outer one is localized on the internal boundary (smallest one) along the defect.}
    \label{fig:nested}
\end{figure*}

\paragraph*{Effective 1D model for edge probability.}
We consider the internal states of the walker that are closest to the edge on the, say, left boundary, i.e., the two states within one lattice site the particle passes if it moves along that edge. The setup is depicted in Fig.~\ref{fig:1D_model}, where all states that belong to the one-dimensional edge are colored in red. For simplicity we will leave out the corner states highlighted in blue. Hence, it suffices to describe the movement along on edge composed of a chain of red states with periodic boundary conditions. Since every lattice side contains two red states, we can describe the dynamics along with the reduced dynamics represented by the graph on the left side of Fig.~\ref{fig:1D_model}. All transitions that leave state $\leftarrow$ or $\downarrow$ and do not enter $\downarrow$  or $\leftarrow$ are considered as scattering towards the bulk represented by black arrows. State $\leftarrow$ can either follow state $\downarrow$  after an internal transition or after an external one.  
The corresponding rate matrix of the dynamics of the probability vector $\bm p_t = (p_{\leftarrow, t}, p_{\downarrow, t})$ along the one-dimensional edge at time step $t$ reads
\begin{equation}
    A = \begin{pmatrix}
        1-D_r & R_1 \\
        R_2 + C_2e^{ik} & 0
        \end{pmatrix}.
    \end{equation}
Note that this matrix does not conserve probabilities due to scatterings into the bulk, i.e., all eigenvalues will have a strictly negative real part. Eigenvalues of $A$ can be computed analytically and read
\begin{equation}
    \lambda_{1,2} = \frac{\text{Tr}(A) \pm \sqrt{[\text{Tr}(A)]^2 - 4 \text{det}(A)}}{2}
\end{equation}
with $\text{Tr}(A) = 1-D_r$ and $\text{det}(A) = -R_1 (R_2 + C_2 e^{ik})$. For $\omega =1$ and $D_r < 0.2$ the eigenspectrum in the complex plane splits into two loops, whereas for $D_r> 0.2$ only one loop is present as can be seen in Fig.~\ref{fig:time_scale_sep}. Here, eigenvalues of the one-dimensional model described above is plotted in black. The gap between the two loops opens around $k = \pi$, i.e., for an antisymmetric phase between the two states. This hints that the antisymmetric coupling between those states is dominated by two different time scales for $D_r < 0.2$. For very small $D_r$ the particle shows a slow dynamics on a lattice site composed of the two internal states and a fast dynamics between neighboring lattice sites.


\newpage
\clearpage

\onecolumngrid 
\section*{Extended figures}

\end{document}